\documentclass[a4paper]{jpconf}
\usepackage{graphicx}
\begin{document}
\title{Modulations in Spectra of Galactic Gamma-ray sources as a result of photon-ALPs mixing.}

\author{Jhilik Majumdar$ ^{1}$, Francesca Calore$ ^{2}$, Dieter Horns$ ^{1}$}

\address{$ ^{1}$Institute for Experimentalphysics, University of Hamburg\\
  $ ^{2}$Laboratoire d'Annecy-le-Vieux de Physique $Th\acute{e}orique$,CNRS \\}

\ead{jhilik.majumdar@desy.de}

\begin{abstract}
Axion like particles (ALPs) are fundamental pseudo scalar particles with properties similar to Axions which are a well-known extensions of the standard model to solve the strong CP problem in Quantum Chromodynamics. ALPs can oscillate into photons and vice versa in the presence of an external tranversal magnetic field. This oscillation of photon and ALPs could have important implications for astronomical observations, i.e. a characteristic energy dependent attenuation in Gamma ray spectra for astrophysical sources. Here we have revisited the opportunity to search Photon-ALPs coupling in the disappearance channel. We use eight years of Fermi Pass 8 data of a selection of promising galactic Gamma-ray source candidates and study the modulation in the spectra in accordance with Photon-ALPs mixing and estimate best fit values of the parameters i.e. Photon-ALPs coupling constant $(g_{\alpha \gamma\gamma} )$ and ALPs mass$ ( m_{\alpha} )$. For the magnetic field we use large scale galactic magnetic field models based on
Faraday rotation measurements and we have also studied the survival probability of photons in the Galactic plane.    \\
\\
\textbf{Key words}: ALPs, Photon, Gamma-ray, Pulsar, Spectrum.
\end{abstract}

\begin{small}
\section*{\centering{Introduction}}

The standard model of particle physics has long been considered to be incomplete because of it's incapability to explain the dark matter problem or the matter-antimatter asymmetry along with the absence of charge parity violation in strong interaction~\cite{Peccei} . As a solution, Peccei and Quinn postulated an additional global symmetry $U(1)_{PQ}$, that is spontaneously broken at some large energy scale $f(a)$. From this $U(1)_{PQ}$ symmetry  a Nambu Goldstone Boson came to existance that was named as Weinberg-Wilczek axion~\cite{Weinberg}. One of the most promising class of particle dark matter candidates are axions~\cite{Barranco}, as they are collisionless, neutral, non-baryonic and may be present in the sufficient quantities to provide the expected dark matter density. ALPs are the valid candidates for cold dark matter Axion-Like particles (ALPs) are more apt and they have the property that they can oscillate into photons or vice-versa in the presence of magnetic fields. An efficient photon-ALPs mixing in the source always means an attenuation in the photon flux~\cite{De Angelis}, whereas the mixing in the intergalactic medium may result in a decrement or enhancement of the photon flux, depending on the distance of the source and the energy considered ~\cite{De Angelis}. Here we investigate this phenomenon in case of six bright gamma-ray pulsar sources and we look for significant spectral irregularities that might be induced by photon-ALPs oscillations in the regular Galactic magnetic field.

\paragraph{Photon-ALPs mixing:} ALPs share same kind of physics to axion: coupling to photon or vice versa in external magnetic field.The equation of the Lagrangian of photon-ALPs coupling describes as:

\begin{equation}
        \mathcal{L} \supset - \frac{1}{4}  g_{\alpha\gamma} F_{\mu\nu} \tilde{F}^{\mu\nu} a = g_{\alpha\gamma\gamma} \vec{E}.\vec{B}a,
\end{equation}

where $a$ is the axion-like field with mass $m_{a}, F_{\mu\nu} $ is the electromagnetic field-strength tensor and $\tilde{F}^{\mu\nu}$ is its dual field , $g_{\alpha\gamma\gamma}$ is the ALPs-photon coupling. Photons, while travelling across the external magnetic field, oscillate with the ALPs state. If the condition, $g_{\alpha\gamma\gamma} Bd \ll 1 $ holds true, the probability of the conversion at a distance $d$ is ~\cite{Mirizzi}: \\

\begin{tiny}
\begin{equation}
        P_{\gamma \rightarrow a}= \frac{g_{\alpha\gamma}^{2}}{8} \left(| \int^d_0 dz'e^{2\pi z'/l_{0}} B_{x}(x,y,z') |^{2}+          
|\int^d_0 dz'e^{2\pi z'/l_{0}} B_{y}(x,y,z') |^{2} \right)\ ,
\end{equation}
\end{tiny}

where $g_{\alpha\gamma\gamma}$ has the dimension of $(Energy)^-1$ and all the parameters are in natural units.

\begin{figure}[h]
\begin{minipage}{16pc}
\includegraphics[width=20pc]{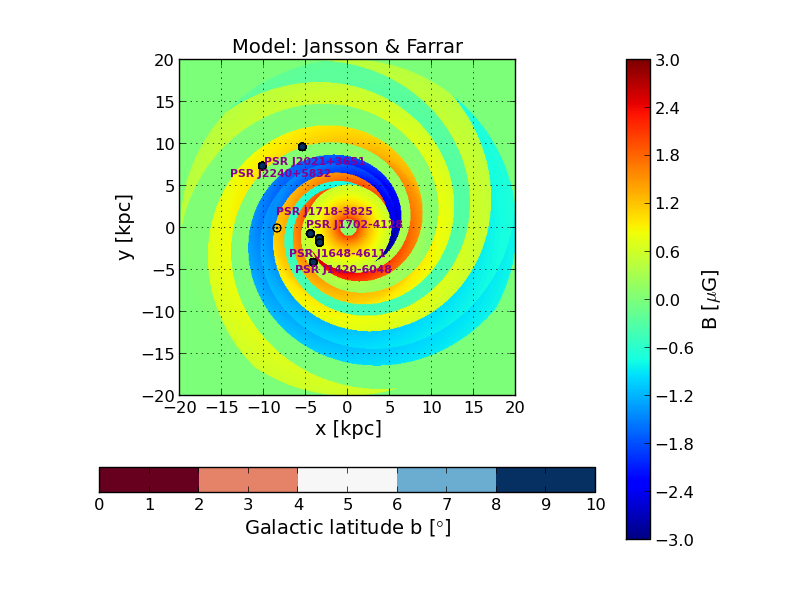}
\caption{\label{label}Source positions in the
plane of Galactic magnetic Field(Jansson $\& $Farrar model).}
\end{minipage}\hspace{4pc}%
\begin{minipage}{16pc}
\includegraphics[width=16pc]{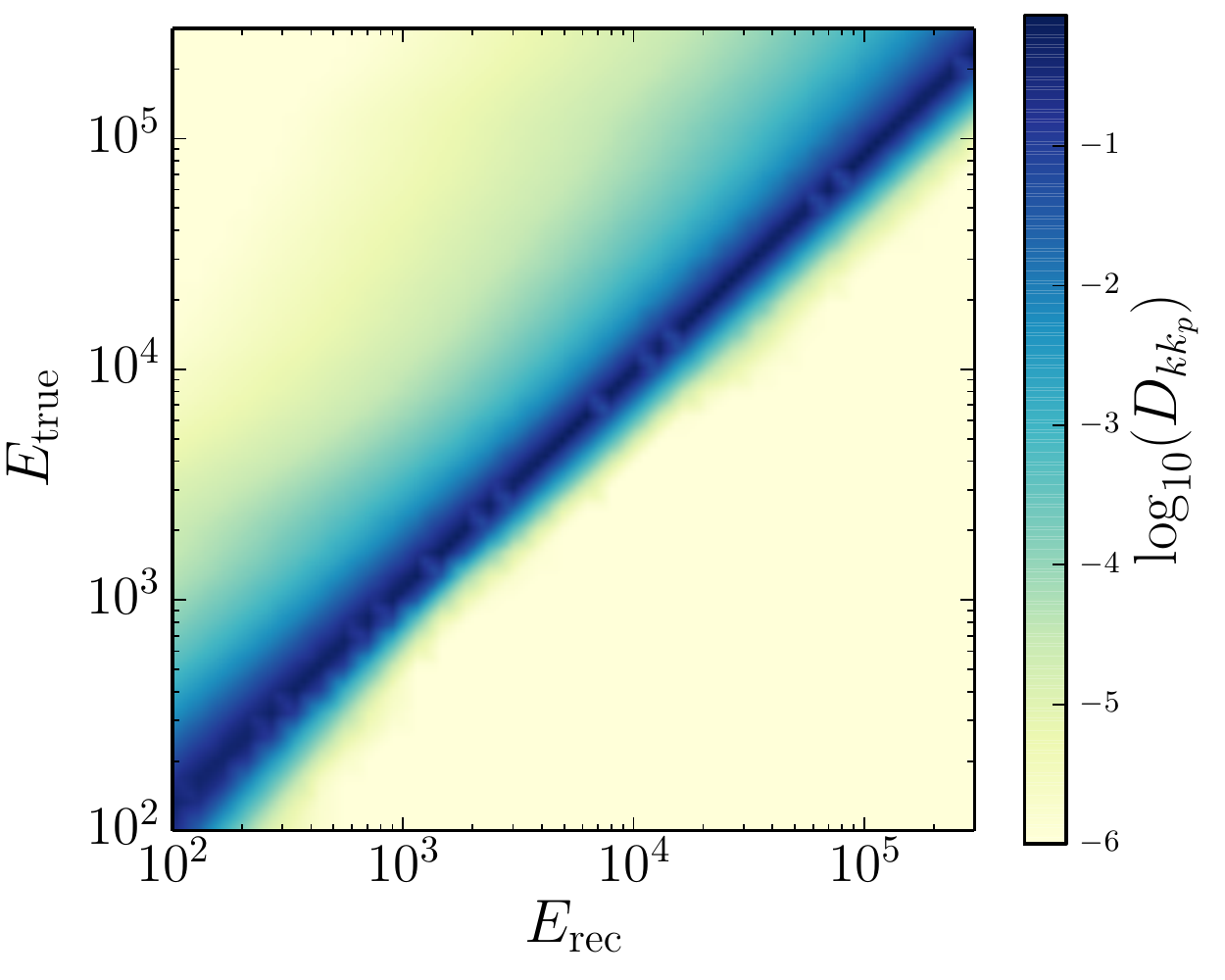}
\caption{\label{label} Energy dispersion matrix derived for all the EDISP event types together.}
\end{minipage} 
\end{figure}

\section*{\centering{Galactic Magnetic Field}}
Best constraints for large scale Galactic Magnetic Field (GMF) are Faraday rotation measures and polarised synchrotron radiation. In our analysis we have taken into consideration one of the most realistic and comprehensive GMF model: Jansson $\&$ Farrar model (2012)~\cite{Jansson}. This  model consists of three components: Disk component, Halo component and Out of plane component.The disk component is partially based on the structure of the NE2001 thermal electron density model~\cite{Sun}. The disk field is constrained to the x-y plane and defined for Galactocentric radius from 3 kpc to 20 kpc with a 'molecular ring' and eight logarithmic spiral regions. The halo field has a purely toroidal component and separate field amplitudes in the north and south. The out of the plane component, formerly known as X component, is to be asymmetric and poloidal. The components of this large scale magnetic field has been updated with the polarized synchrotron and dust emission data measured by Planck satellite~\cite{Planck}.   

\section*{\centering{Source selection}}

\paragraph{Fermi-LAT observations of gamma-ray pulsars:}
In the present work we use gamma-ray data from the Fermi-LAT, a pair conversion telescope collecting gamma rays between 20 MeV to more than 300 GeV . To date, about 160 gamma-ray pulsars have been discovered with Fermi-LAT \cite{Fermi}. We have chosen a list of  six bright gamma-ray  pulsars  named as : PSR J2021+3651 , PSR J1420-6048, PSR J2240+5831, PSR J1648-4611, PSR J1718-3825  and PSR  J1702-4182. Photons coming from these sources, while traversing towards the observer, have to cross the spiral arms. That is the main criteria to choose these sources. PSR J2021+3651 is 17 kyr old rotation powered pulsar detected in radio, X-rays, and gamma rays and it is quite similar  to the Vela pulsar ~\cite{Kirichenko} based on  the optical upper limits.  PSR J1420-6048 is a 68 ms pulsar surrounded by a Nebula and it has been observed in X-ray, radio and infrared ~\cite{J1420-6048}.  PSR J1648-4611 is a very-high-energy (VHE) $\gamma - ray$ source obverved by HESS ~\cite{Abramowski} as well as Fermi LAT, which is is centered on the massive stellar cluster Westerlund 1. PSR J1718-3825 and  PSR  J1702-4182  has been discovered associated wth  the pulsar wind nebula  candidate: HESS J1718-385  ~\cite{J1718-3825} and HESS J1702-420 ~\cite{J1702-4182} with a spin-down age of 55 kyr respectively in 2006 . All of these bright six pulsars have low galactic latitude so that the emitted  photons can directly penetrate into the Galactic spiral arms (see Fig.~1).   In order to estimate systematic uncertainties on the observed spectrum we use as a reference the Vela pulsar~\cite{Ackermann}. This pulsar is very close; the spectrum is very well measured and does not show any spectral distortion. We use the same technique that has been done by the Fermi Collaboration to derive the systematics using Fermi-LAT Pass 7 data~\cite{Ackermann}.

\begin{figure}
\setlength{\unitlength}{.8cm}
\begin{center}
\begin{picture}(15,4.5)
 \put(-2.5,0){\includegraphics[width=5.5cm]{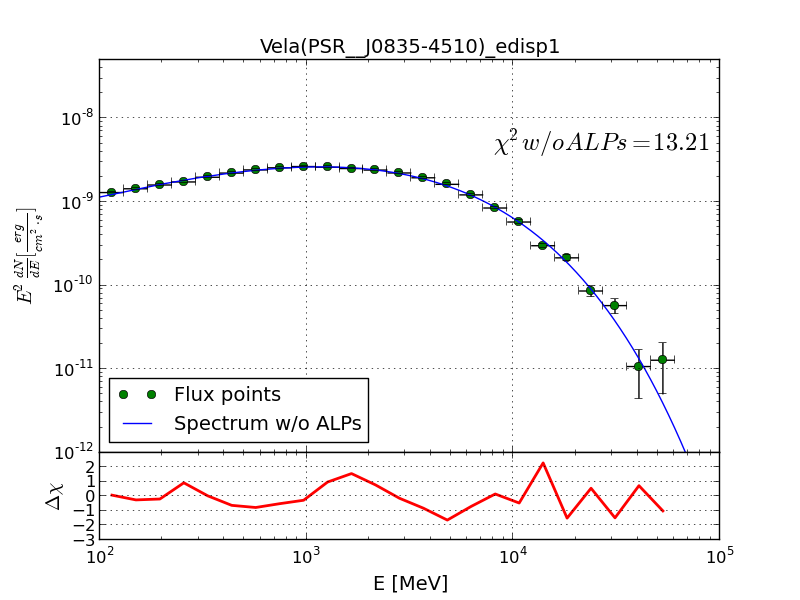}}
 \put(3.8,0){\includegraphics[width=5.5cm]{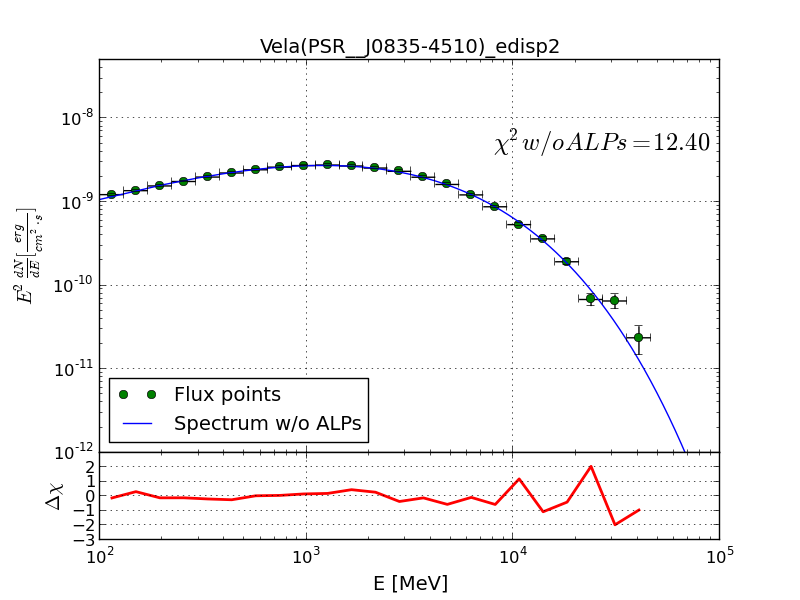}}
 \put(10.2,0){\includegraphics[width=5.5cm]{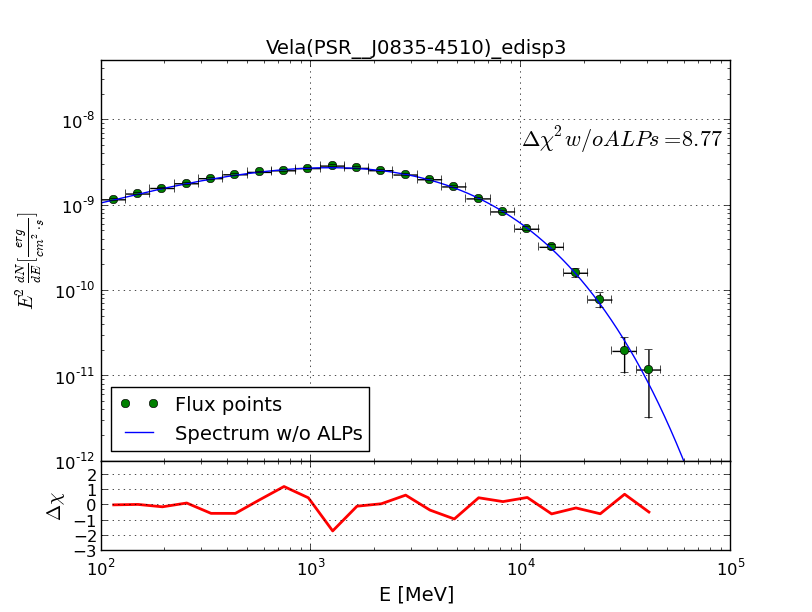}}
 \end{picture}
\end{center}
\caption{Systematic uncertainties of Vela Pulsar for EDISP1, EDISP2,EDISP3 event types.}
\end{figure} 

\section*{\centering{Analysis}}
 
We use eight and half  years of Fermi-LAT Pass 8 data with P8R2 SOURCE V6 IRFs of six bright pulsar candidates, i.e. PSR J2021+3651 et. al and Vela. Pass 8 data has  an  improved  angular  resolution,  a
broader energy range, larger effective area, as well as reduced  uncertainties  in  the  instrumental  response  functions ~\cite{Ackermann}. For spectral modelling of Fermi-LAT sources Enrico binned likelihood optimization technique\cite{Enrico} is performed for 25 energy bins.  
All of the pulsar spectrum is modelled by a power law with  exponential cutoff:
  \begin{equation}
 \frac{dN}{dE}= N_{0} \left(\frac{E}{E_{0}}\right)^{-\Gamma} \exp\left(-\frac{E}{E_{\rm cut}}\right)
 \end{equation}

For Vela we use a power law with super exponential cutoff :
 \begin{equation}
 \frac{dN}{dE}= N_{0} \left(\frac{E}{E_{0}}\right)^{-\Gamma_{1}} \exp\left[\left(-\frac{E}{E_{\rm cut}}\right)^{\Gamma_{2}}\right]
 \end{equation}

We perform a fit to the data, minimising the $\chi^{2}$ function~\cite{Ackermann1}~\cite{Jogler} .
We have checked that the log(likelihood) has a parabolic pattern and thus a  $\chi^{2}$ analysis is appropriate. We derive the energy dispersion matrix for one energy dispersion event type (EDISP) via the transformation of the number of counts in true energy of a particular energy bin to the number of counts in that bin of reconstructed energy (see Fig.~2) and we fully take it into account in the fit. We investigate the signature of photon-ALPs oscillations, including the effect of oscillations in the predicted spectra:
  
\begin{equation}
\left(\frac{dN}{dE}\right)_{\rm fit} =D_{kk_{p}} . P_{\gamma \rightarrow a}\left( E,g_{a\gamma\gamma},m_{a},d\right).\left(\frac{dN}{dE}\right)
\end{equation}


\paragraph{Systematic uncertainties of Vela:} For P8R2\_SOURCE\_V6 event class, systematic uncertainties in effective area are derived to be about 2.4 \% for EDISP1, EDISP2 ,  EDISP3 event types for Vela considering the energy range from 100 MeV to 300 GeV(see Fig.~3). We calculated the systematics in such a way so that the $\chi^2$ per Degrees of Freedom(dof) we get $\sim$ 1 and it's an acceptable fit.

\begin{figure}
\setlength{\unitlength}{.9cm}
\begin{center}
\begin{picture}(12,4)
 \put(-2,0){\includegraphics[width=7.5cm]{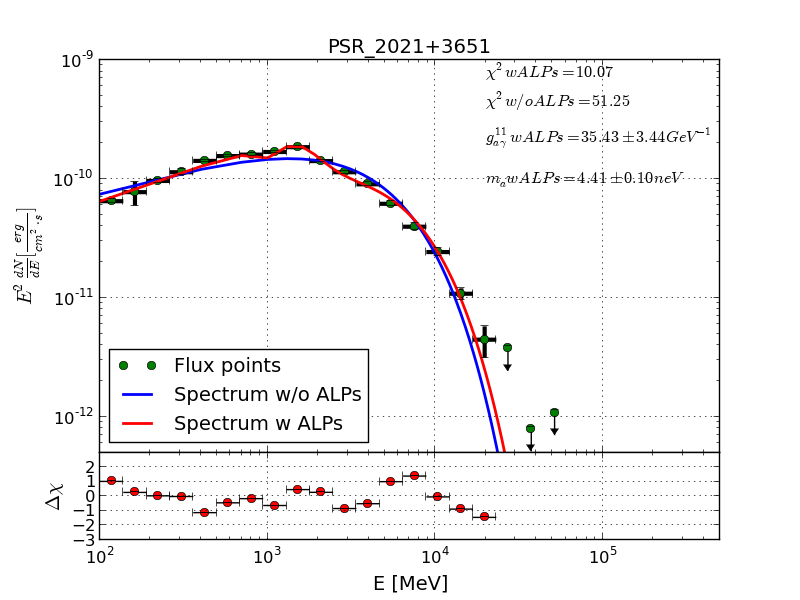}}
 \put(6,0){\includegraphics[width=7.5cm]{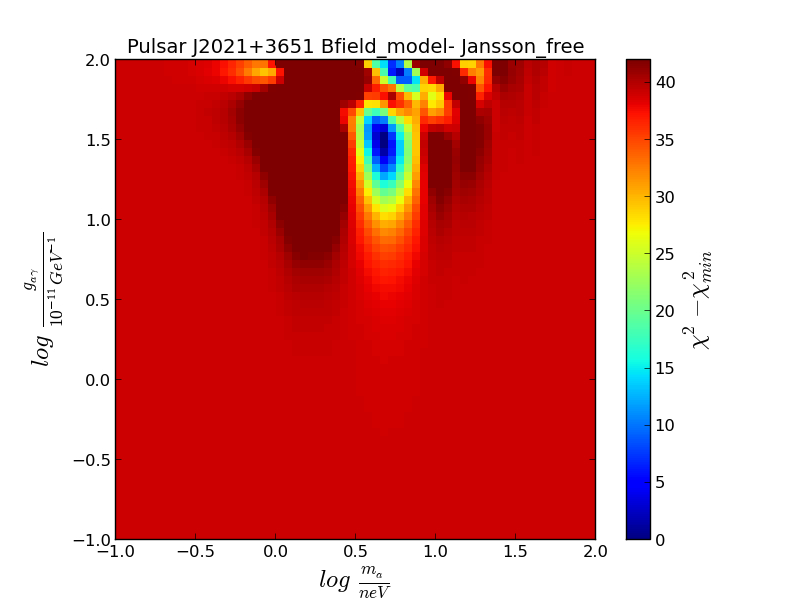}}
 \end{picture}
\end{center}
\caption{Left panel: Best-fit model of the spectrum of PSR J2021 + 3651. Right panel: The $ \chi^{2}$ scan as function of photon-ALPs coupling and ALPs mass. }

\end{figure}

\begin{figure}
\setlength{\unitlength}{.9cm}
\begin{center}
\begin{picture}(10,6.8)
 \put(-3,0){\includegraphics[width=7.5cm]{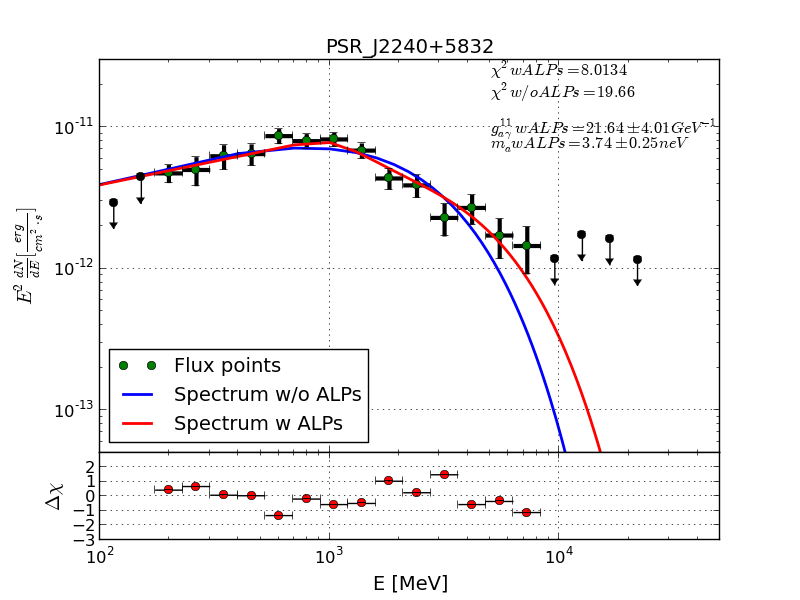}}
 \put(5,0){\includegraphics[width=7.5cm]{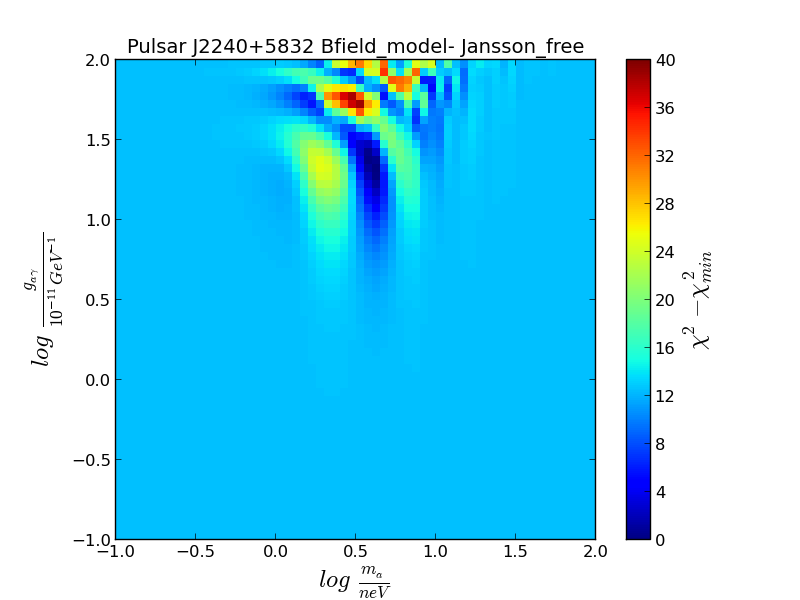}}
 \end{picture}
\end{center}
\caption{Left panel:  Best-fit model of the spectrum of PSR J2240+5831. Right panel: The $ \chi ^{2}$ scan as function of photon-ALPs coupling and ALPs mass.}

\end{figure}

\section*{\centering{Results and Discussion}}

Apparent absorption in the energy spectrum, i.e. disappearance channel, is expected due to the mixing of photon and pseudo-scalars in the Galactic magnetic field and depends on the photon-ALPs coupling and the mass of the ALPs. On the other hand, this spectral feature of disappearance channel depends also on the transversal magnetic field along the line of sight and the distance to the source. For allowed values of $g_{a\gamma} \sim  10^{-11} \rm GeV^{-1} $ the mixing is non-linear in the spiral arms and in the large scale field of the inner Galaxy. Including the effect of spectral modulation due to photon-ALPs oscillations, leads to a significant improvement for PSR J2021+3651 (see Fig.~4) and PSR J1420-6048 (see Fig.~5). We have derived the values for couplings in the range 10 -- 60 $ \times 10^{-11} \rm GeV^{-1} $ and ALPs masses 3 -- 8 neV. In the plots we show that we get a significant improvement in the $\chi^2$ of about 51 to 10 for the pulsar candidate PSR J2021+3651 as we introduce two additional mixing parameters. We can say the fit results   with ALPs parameters is a good fit as the $\chi^2$ per dof is $\sim$ 1 for all the pulsar spectrum. The best fit $\chi^2$ we get for our candidate spectrum  In the fits we also include systematic uncertainties as derived from the Vela pulsar analysis. In the photon-ALPs parameter space plot we clearly see the blue stripes that are created for the photon-ALPs oscillation in the Galactic magnetic field. We make a combined photon-ALPs parameter space plot out of our all six pulsars(see Fig.~6). Best parameter space we get from the combined plot for $g_{a\gamma}\times 10^{-11} \rm GeV^{-1} $ from 6 --26  and mass around 2--4 neV. The significance of our results has been estimated 6.3 $\sigma$.

\begin{figure}
\setlength{\unitlength}{.9cm}
\begin{center}
\begin{picture}(10,6.8)
 \put(2,0){\includegraphics[width=7.5cm]{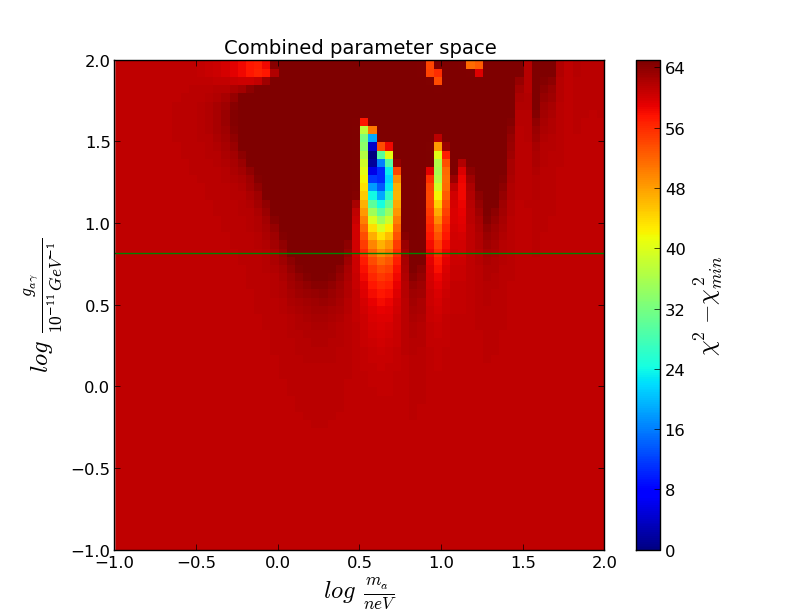}}
  \end{picture}
\end{center}
\caption{Combined $ \chi^{2}$ scan as function of photon-ALPs coupling and ALPs mass.}

\end{figure} 

The upper bound of has been set by the CAST experiment as $g_{a\gamma} <6.6 \times 10^{-11} \rm GeV^{-1} $ (dotted line in Fig.6)~\cite{CAST} and our best fit parameter space is coinciding with the excluded region. The Photon-Alps mixing is very much magnetic field model dependent  and if we include uncertainties in galactic spiral arms, which varies in kpc range, the whole parameter space go down and submerge with the space left by the obdervation of NGC 1275 by Fermi LAT~\cite{Ajello}. We will revisit the same analysis for other gamma-ray sources in future.

\end{small}
        


\end{document}